\documentstyle[12pt]{article}
\begin{document} \baselineskip 28pt

\font\bbb = msbm10
\def\Bbb#1{\hbox{\bbb #1}}
\font \lbb = msbm7
\def\LBbb#1{\hbox{\lbbb #1}}

\newcommand{\BB} {{\Bbb C }}
\newcommand{\CC}   {{\Bbb C}}
\newcommand{\RR}  {{\Bbb R}}
\newcommand{\ZZ} {{\Bbb Z}}
\newcommand{\EE} {{\Bbb E}}
\newcommand{\NN} {{\Bbb N}}

\begin{center}
{\large\bf The Correlations of Greenberger-Horne-Zeilinger States
Described by Hilbert-Schmidt Decomposition}\\ \ \\
{\bf Y. Ben-Aryeh and A. Mann}\\
{\em Department of Physics, Technion-Israel Institute of
Technology\\Haifa 32000, Israel}\\ \ \\
\end{center}

Using simple quantum analysis we describe the correlations of
Greenberger-Horne-Zeilinger (GHZ) states by the use of
Hilbert-Schmidt (HS) representation. Our conclusion is that while
these states disprove local-realism they do not prove any
nonlocality property.

\pagebreak

\noindent{\bf 1. INTRODUCTION}\\

Greenberger-Horne-Zeilinger (GHZ) states have attracted much
attention in recent literature [1-11] as experiments which are
done with these states disprove local-realism, without the use of
Bell's inequalities. In many articles it was claimed that GHZ
states violate locality. We would like to quote here some
examples: In Ref. 3 ``GHZ states have been fascinating quantum
systems to reveal {\em the nonlocality of the quantum world}''
(our emphasis). In Ref. 7 (in the abstract): ``A scheme is
proposed for generating maximally entangled GHZ atomic states {\em
for testing quantum nonlocality}'' (our emphasis). In Ref. 8 (in
the abstract): ``enable various novel tests of {\em quantum
nonlocality}'' (our emphasis). In Ref. 10, in Caption to Figure 1:
``GHZ tests of quantum nonlocality". In Ref. 11 (in the abstract):
``We propose an experimentally feasible scheme to {\em demonstrate
quantum nonlocality}'' (our emphasis). In Ref. 12 (in the
abstract) ``it is possible to {\em demonstrate nonlocality} for
two particles without using inequalities''. These expressions and
many more which can be found in the usual literature lead to the
impression that the quantum world is nonlocal.  In the present
Letter we would like to analyze the correlations obtained for GHZ
states and show that ``locality'' is not violated by these
correlations. We adopt here the definition given in Ref.~12: ``The
assumption of locality is that the choice of measurement on one
side cannot influence the outcome of any measurement on the other
side.'' It is obvious that once observer ``{\em a}'' does anything
to his part of the system, measurement on the ``{\em a}'' system
will give a new result, and the correlations between {\it a} and
the rest of the world will change. This does not contradict
locality, and is obviously true also classically. However,
locality implies that no matter what is done in the ``{\em a}''
system, it should not affect the measurements on {\it b} and {\it
c} and {\em also} not affect the correlations between {\it b} and
{\it c}.  By following this definition we show in the following
pure quantum mechanical (QM) analysis that experiments with GHZ
states do not violate locality. Although we treat here a specific
system, by following the present approach a similar QM analysis
can be done also for other entangled systems.\\

\noindent{\bf 2. ANALYSIS}

Greenberger, Horne, Shimony and Zeilinger [1] have suggested a
gedanken three-particle interferometer which has been described
[1] as follows: The source emits a triple of particles 1, 2, and
3, in six beams, with the state given by
\begin{eqnarray}
| \psi \rangle = (1/\sqrt{2} ) \left[ |a\rangle_1 |b\rangle_2
|c\rangle_3 + |a'\rangle_1 |b'\rangle_2 |c'\rangle_3 \right] \ . 
\end{eqnarray}

The three particles 1, 2, and 3 emerge either through $a$, $b$,
and $c$ apertures or through $a'$, $b'$, and $c'$, respectively. A
phase shift $\phi_1$ is imparted to beam $a'$ of particle 1, and
beams $a$ and $a'$ are brought together on a beam-splitter before
illuminating detectors $d$ and $d'$. Likewise for particles 2 and
3, with their respective apertures, phase shifts and detectors.
The evolution of the kets $|a\rangle_1$ and $|a'\rangle_1$ is
given by
$$|a\rangle_1 \rightarrow (1/\sqrt{2}) \left[|d\rangle_1 + i|d'\rangle_1
\right. \eqno(2-a)$$
and
$$ |a'\rangle_1 \rightarrow (1/\sqrt{2}) e^{i\phi_1}
\left[|d'\rangle_1 + i|d\rangle_1\right] \ , \eqno(2-b)$$ where
the ket $|d\rangle_1$ denotes particle 1 directed toward detector
$d_1$, etc. The particle 2 beams and the particle 3 beams are
subjected to similar treatment and hence undergo similar
evolutions. A state with eight terms develops from which we obtain
amplitudes and hence probabilities of detection of the three
particles by the triple detectors ($d, \ e, \ f$), the triple of
detectors $(d'$, $e$, $f$), etc. (An analysis of possible
experiments which can be done on this system which refute
local-realism is described in Ref. 1).

The quantum state given by Eq.~(1) can be considered formally as a
three spin-$\frac{1}{2}$ system (denoted by $a,$ $b$, and $c$).
$|a\rangle$ and $|a'\rangle$ may be represented by the two levels
$\left( \begin{array}{c} 1 \\ 0 \end{array}\right)$ and $\left(
\begin{array}{c} 0 \\ 1 \end{array}\right)$ of the first
spin-$\frac{1}{2}$ system, $|b\rangle$ and $|b'\rangle$ are the
two levels of the second spin-$\frac{1}{2}$ system, etc.

In this representation the density matrix $|\psi \rangle \langle
\psi|$ corresponding to the state $|\psi\rangle$ of Eq.~(1) is
given by:
$$
\rho = \left(
\begin{array}{cccccccc} \frac{1}{2} & 0 & 0 & 0 & 0 & 0 & 0 &
\frac{1}{2} \\[0.15cm]
0 & 0 & 0 & 0 & 0 & 0 & 0 & 0 \\[0.15cm]
0 & 0 & 0 & 0 & 0 & 0 & 0 & 0 \\[0.15cm]
0 & 0 & 0 & 0 & 0 & 0 & 0 & 0 \\[0.15cm]
0 & 0 & 0 & 0 & 0 & 0 & 0 & 0 \\[0.15cm]
0 & 0 & 0 & 0 & 0 & 0 & 0 & 0 \\[0.15cm]
0 & 0 & 0 & 0 & 0 & 0 & 0 & 0 \\[0.15cm]
\frac{1}{2} & 0 & 0 & 0 & 0 & 0 & 0 & \frac{1}{2} \end{array}
\right) \ . \eqno(3) $$

While this form of the density matrix seems quite simple, the
locality of GHZ states is demonstrated in a better way by using
the HS  representation [13,14] of this density matrix.

For an entangled state of three two-level particles (denoted by
$a, \ b, \ c$), the HS decomposition becomes [14]:

$$ 8\rho = \begin{array}[t]{l} (I)_a \bigotimes (I)_b \bigotimes (I)_c
+ (\vec{r}\cdot\vec{\sigma})_a \bigotimes(I)_b \bigotimes (I)_c +
(I)_a \bigotimes (\vec{s} \cdot \vec{\sigma})_b \bigotimes (I)_c
\\[0.15cm]
+ (I)_a \bigotimes (I)_b \bigotimes (\vec{p}\cdot\vec{\sigma})_c +
\sum_{mn} t_{mn} (I)_a \bigotimes(\sigma_m)_b \bigotimes
(\sigma_n)_c\\[0.15cm]
+\sum_{k\ell} o_{k\ell}(\sigma_k)_a \bigotimes (I)_b \bigotimes
(\sigma_\ell)_c + \sum_{ij} p_{ij} (\sigma_i)_a
\bigotimes(\sigma_j)_b \bigotimes (I)_c \\[0.15cm]
+ \sum_{\alpha,\beta,\gamma}
R_{\alpha\beta\gamma}(\sigma_\alpha)_a \bigotimes(\sigma_\beta)_b
\bigotimes (\sigma_\gamma)_c  \ . \end{array} \eqno(4)$$

Here $I$ stands for the unit operator, $\vec{r}, \ \vec{s}$, and
$\vec{p}$ belong to  $\RR^3, \ {\sigma_n} \ (n= 1,2,3)$ are the
standard Pauli matrices. The coefficients $t_{mn}$, $o_{k\ell}$,
and $p_{ij}$ form real $3 \times 3$ matrices.

The coefficients $R_{\alpha\beta\gamma}$ form a real $3 \times 3
\times 3$ tensor, related to the density matrix $\rho$ by
$$R_{\alpha\beta\gamma} = \left(\frac{1}{8}\right) Tr
\left[\rho(\sigma_\gamma)_c \bigotimes (\sigma_\beta)_b \bigotimes
(\sigma_\alpha)_a\right] \ . \eqno(5)$$ The coefficients $t_{mn}$
are related to the density matrix $\rho$ by
$$t_{mn} =\left(\frac{1}{8}\right) Tr \left[\rho(\sigma_n)_c \bigotimes (\sigma_m)_b
\bigotimes (I)_a\right] \ , \eqno(6)$$ and similar relations hold
between other coefficients and the density matrix. In deriving
such relations we use the simple relation
$Tr[\sigma_i\sigma_j]=2\delta_{ij}$. We find that the general
entangled state of three two-level systems is described by 63
parameters: 9 for $\vec{r}, \ \vec{s}$ and $\vec{p}$, 27 for
$t_{mn}$, $o_{k\ell}$ and $p_{ij}$ and 27 for
$R_{\alpha\beta\gamma}$. The parameters $\vec{r}, \ \vec{s}$ and
$\vec{p}$ can be obtained from measurements on one arm of the
measurement device, $t_{mn}$, $o_{k\ell}$ and $p_{ij}$ can be
obtained from measurements on the corresponding two arms of the
measurement device [e.g., by using Eq.~(6)] and
$R_{\alpha\beta\gamma}$ can be obtained from measurements on the
three arms of the measuring device, [e.g., by using Eq. (5)].
Local realism has been refuted by applying different sets of
measurements in the different arms of the measurement device [1].
Although the representation (4) assumes axes of measurements $x, \
y, \ z$ of a certain basis $F$ corresponding to $\sigma_1, \
\sigma_2, \sigma_3$, changes of axes of measurement can be
obtained by rotation from the basis $F$ to another $F'$ [13-14]:
$$(\vec{\sigma}^{F'})_a = O_1(\vec{\sigma}^F)_a; \
(\vec{\sigma}^{F'})_b = O_2(\vec{\sigma}^{F})_b;
(\vec{\sigma}^{F'})_c = O_3(\vec{\sigma}^{F})_c \ . \ \eqno(7) $$

The 63 parameters defining the density matrix can be obtained by
measurements in the $x,y$ and $z$ directions. If the measurements
are done along different axes (e.g., $x', y', z'$) one should
transform these parameters accordingly. The essential point here
is that the rotation of axes in system $a$ can be done {\it
independently} of the rotation of axes in system $b$ or $c$ so
that in addition to the 63 parameters which have been fixed by
local interaction in the past, each observer can rotate
individually his axis of measurement. The correlations are, of
course, changed but only due to {\em local} operations.

The HS representation for the density matrix (3) corresponding to
the wavefunction $|\psi\rangle$ of Eq.~(1) has been already
evaluated in our previous article [14]. The parameters for this
state are given by $$R_{122} = R_{212} = R_{221}= -1; t_{33} =
o_{33} = p_{33} = R_{111} = 1 \ ,  \eqno(8)$$ and all other
parameters are equal to zero.

Experimentally the states $|d\rangle_1$ and $|d'\rangle_1$ of
Eqs.~(2) are obtained from the states $|a\rangle_1$ and
$|a'\rangle_1$ by the use of 50-50 beam-splitter transformation.
This transformation can be described by the following unitary
transformation:
$$\left(\begin{array}{c} d\\d'\end{array}\right) =
\frac{1}{\sqrt{2}} \left(\begin{array}{cc} 1 & -ie^{-i\phi_1}
\\-i & e^{-i\phi_1}\end{array}\right)
\left(\begin{array}{c} a\\a'\end{array}\right) = U_1
\left(\begin{array}{c} a\\a'\end{array}\right) \eqno(9) $$ The
operation of the beam-splitter on side $a$ leads to a change in
the HS representation of Eq. 4 by changing $(\sigma_i)_a$
$(i=1,2,3)$ everywhere in this equation, into $U_1(\sigma_i)_a
U^\dagger_1$, where $U_1$ is given by Eq.~(9). Here again the
beam-splitter transformation on side $a$ does not affect other
sides of the system as given by the HS representation.  The HS
decomposition shows this very simply. The correlations between
system {\it a } and systems {\it b} and {\it c} are changed by
changing $(\sigma_k)_a$, $(\sigma_i)_a$ and $(\sigma_\alpha)_a$ in
the 6'th, 7'th and 8'th terms of Eq. (4), respectively. The
correlations between {\it b} and {\it c} given by the 5'th term of
Eq. (4) and the coefficients $t_{mn}$ are unaffected by the
beam-splitter or any other  interaction employed by ``{\it a}''.
In a similar way, one can see the local effects of the
beam-splitters in other sides of the GHZ system. The correlations
are, of course, changed due to the beam-splitter transformation
but only due to {\em local} operations. When these occur in
subsystem ``{\it a}'', measurements of ``{\it a}'' and
correlations with ``{\it a}'' are affected, of course. However,
all other correlations are not affected, which is the essence of
locality.

In conclusion, locality, as defined in the present letter, is not
violated by QM. EPR correlations are fixed by the local
interaction which occurred in the past.~Local-realism is, however,
refuted. \pagebreak

 \noindent{\bf References}
\begin{enumerate}
\item D.M. Greenberger, M.A. Horne, and A. Zeilinger, in {\em
Bell's Theorem, Quantum Theory, and Conceptions of the Universe},
edited by M. Kafatos (Kluwer, Dordrecht, 1989). D.M. Greenberger,
M.A. Horne, A. Shimony, and A. Zeilinger, {\em Am. J. Phys.} {\bf
58}, 1131 (1990).
\item A. Zeilinger, M.A. Horne, H. Weinfurter and M. Zukowski,
{\em Phys. Rev. Lett.} {\bf 78}, 3031 (1997).
\item J.-W. Pan and A. Zeilinger, {\em Phys. Rev.} A. {\bf 57}, 2208 (1998).
\item D. Bouwmeester, J.-W. Pan, M. Daniell, H. Weinfurter, and A.
Zeilinger, {\em Phys. Rev. Lett.} {\bf 82}, 1345 (1999).
\item M. Zukowski, A. Zeilinger, M.A. Horne and A. Weinfurter,
{\em Int. Jour. Theor. Phys.} {\bf 38}, 501 (1999).
\item S.-B. Zheng, {\em Quant. Semiclass. Opt.} {\bf 10}, 695 (1998).
\item S.-B. Zheng, {\em Phys. Rev. Lett.} {\bf 87}, 230404 (2001).
\item J.-W. Pan, M. Daniell, S. Gasparoni, G. Weihs and A.
Zeilinger, {\em Phys. Rev. Lett.} {\bf 86}, 4435 (2001).
\item J.-W. Pan, D. Bouwmeester, M. Daniell,  H. Weinfurter and A.
Zeilinger, {\em Nature} {\bf 403}, 515 (2000).
\item T.E. Keller, M.H. Rubin and Y.H. Shih, {\em Fortschr.\ Phys.} \ {\bf
46}, 673 (1998).
\item X.-B. Zhou, Y.-S. Zhang and G.-C. Guo, {\em J. Opt. B: Quant.
Semiclass. Opt.} {\bf 5}, 164 (2003).
\item L. Hardy, {\em Phys. Rev. Lett.} {\bf 71}, 1665 (1993).
\item R. Horodecki and P. Horodecki, {\em Phys. Lett.} A {\bf 210}, 227
(1996).
\item Y. Ben-Aryeh, A. Mann, and B.C. Sanders, {\em Found. Phys.} {\bf 29},
1963 (1999).
\end{enumerate}

\end{document}